\documentstyle[12pt]{article}
\def\del        {  \partial  }
\def\half       {  {1\over 2}  }

\def\defint#1#2 {  \int_{#1}^{#2}  }
\def\rootof#1   {  \left( #1 \right)^{1/2}  }
\def\deldel#1   {  {\partial\over \partial #1}  }

\def\abs#1      {  \vert #1 \vert  }
\def\ie         {  {\it i.e.}      }
\def\evalat#1   {  \left\vert_{#1} \right. }
\def\e          { {\rm e}  }

\def\lsim    {\lower .65ex \hbox{\ $\stackrel{<}{\sim}$\ } }
\def\gsim    {\lower .65ex \hbox{\ $\stackrel{>}{\sim}$\ } }

\def\calF       { {\cal F} }

\def\calU       { {\cal U} }

\def\vecii#1#2      {  \left(\begin{array}{c}#1\\#2\end{array}\right)  }
\def\veciii#1#2#3   {  \left(\begin{array}{c}#1\\#2\\#3\end{array}\right)  }
\def\veciv#1#2#3#4  {  \left(\begin{array}{c}#1\\#2\\#3\\#4
                                 \end{array}\right)  }
\def\vecv#1#2#3#4#5 {  \left(\begin{array}{c}#1\\#2\\#3\\#4\\#5
                                 \end{array}\right)  }

\def\matrixii#1#2#3#4            {  \left(\begin{array}{cc}#1&#2\\#3&#4
                                       \end{array}\right) }
\def\matrixiii#1#2#3#4#5#6#7#8#9 {  \left(\begin{array}{ccc}#1&#2&#3\\
                                     #4&#5&#6\\#7&#8&#9\end{array}\right)  }
\def\mativ#1#2#3#4               {  \left(\begin{array}{cccc}
                                       #1\\#2\\#3\\#4\end{array}\right) }
\def\matv#1#2#3#4#5              {  \left(\begin{array}{ccccc}
                                     #1\\#2\\#3\\#4\\#5\end{array}\right)  }
\def\eqabegin         {  \begin{eqnarray}  }
\def\eqaend           {  \end{eqnarray}  }
\def\nn               {  \nonumber  }
\def\bracetwo#1#2     {  \left\{ \begin{array}{l} #1 \\ #2 \end{array}
                         \right.  }
\def\bracetwocases#1#2#3#4  {   \left\{ \begin{array}{ll} #1 &
                                 \qquad #2 \\
                                 #3 & \qquad #4 \end{array} \right.  }
\def\bracebegin#1     {  \left\{ \begin{array}{#1}   }
\def\braceend         {  \end{array}\right.   }
\def\parn              {  \par\noindent }
\def\parbigskip        {  \par\bigskip  }
\def\parmedskip        {  \par\medskip  }
\def\parsmallskip      {  \par\smallskip  }
\def\parbigskipn        {  \par\bigskip\noindent  }

\def\parag#1           {\paragraph{#1} \mbox{ }\parmedskip\noindent}
\def\boxit#1#2      {  \vbox{\hrule\hbox{ \hskip -4.1pt \vrule\kern3pt
                       \vbox
                    {  \hsize #1 \strut\kern3pt #2 \kern3pt\strut  }
                       \kern3pt  \vrule} \hrule  } }
\def\centerbox#1#2  {  \mbox{  }\par\bigskip  \hfil \boxit{#1}{#2} \hfil
                       \par\bigskip\noindent }
\def\rightbox#1#2   {  \hfill\boxit{#1}{#2}  }
\def\leftbox#1#2    {  \boxit{#1}{#2}  }
\def\fullbox#1      {  \boxit{\textwidth}{#1}  }
\def\trianglemap#1#2#3#4#5#6  {   {\large $$ \begin{array}{rcl} #1\!\!\!
                                  &{\stackrel{{\scriptstyle #2}}{
                              \longrightarrow   }}&\!\!\!  #3 \\
                            { } & {\scriptstyle #4}\!\!\!\searrow \quad
                                \swarrow \!\!\!{\scriptstyle #5}& { } \\
                                  { } & #6 & { } \end{array} $$ }    }
\def\squaremap#1#2#3#4#5#6#7#8    { {\large $$ \begin{array}{ccc}#1 &
                   \stackrel{{\scriptstyle #2}}{\longrightarrow} & #3 \\
                     {\scriptstyle #4}\!\downarrow & { } & \downarrow \!
                     {\scriptstyle #5}\\ #6 &\!\!
                      \longrightarrow_{{ }_{\!\!\!\!\!\!\!\!\!\!\!
                      {\scriptstyle #7}}}    &#8 \end{array} $$ }   }
\def\righttrianglemap#1#2#3#4#5#6  {  {\large $$ \begin{array}{rcl}
                 #1\!\! & \stackrel{{\scriptstyle #2}}{\longrightarrow}
                      & #3 \\  { }&\!\!{\scriptstyle #4}\!\!\searrow
                      & \downarrow \!\!{\scriptstyle #5}\\
                      { }&{ }& #6 \end{array} $$ }   }
\def\rightfigspacebegin  {  \par\noindent\begin{minipage}[t]{10cm}  }
\def\rightfigspaceend    {  \end{minipage}\par\noindent  }
\def\leftfigspacebegin   {  \par\noindent
                             \hspace*{10cm}\begin{minipage}[t]{6cm} }
\def\leftfigspaceend     {  \end{minipage}\par\noindent  }
\def\titleandfile#1#2   {  \begin{center}{\Large\bf #1}\end{center}
                            \par\begin{flushright} #2 \end{flushright}  }
\def\msection#1      {  \begin{center} \section{#1} \end{center}   }
\def\nsection#1      {  \let\boldface\bf \def\bf{} \section{#1}
                           \let\bf\boldface   }
\def\mnsection#1     {  \begin{center} \nsection{#1} \end{center}  }
\def\capsection#1    {  \let\boldface\bf \def\bf{\sc} \section{#1}
                           \let\bf\boldface   }
\def\mcapsection#1   {  \begin{center} \capsection{#1} \end{center} }
\def\sectionnumbering { \setcounter{equation}{0}
         \renewcommand{\theequation}{\arabic{section}.\arabic{equation}}}
\newcommand{\nullify}[1]{}
\def\period{\, .}
\def\comma{\, ,}
\def\quarter{{1\over 4}}

\def\overrttwo{{1 \over \sqrt{2}}}
\def\overi{{1\over i}}
\def\ff{\gamma^2}  
\def\xplus{{x^+}}
\def\xminus{{x^-}}
\def\yplus{{y^+}}
\def\yminus{{y^-}}
\def\delplus{\del_+}
\def\delminus{\del_-}
\def\delsigma{\del_\sigma}
\def\deltamn{\delta_{m+n,0}}
\def\ket#1{\mid #1 >}

\def\rhodot{\dot{\rho}}

\def\Phidot{\dot{\Phi}}

\def\ghat{\hat{g}}

\def\calF{{\cal F}}
\def\calUinv{\calU^{-1}}

\def\deltamn{\delta_{m+n,0}}

\def\downvacket{\mid 0 >_\downarrow}
\def\pdownvac{\mid \vec{P} >_\downarrow}

 \def\phione{\phi_1} \def\phitwo{\phi_2}
        \def\Ldl{L^{dL}}
            \def\Lf{L^f}

\def\Pplus{P^+} \def\Pminus{P^-}

\def\vecpf{\vec{p}_f}
\def\vecal{\vec{\alpha}}
\def\dhat{\hat{d}}
\def\dzerohat{\hat{d}_0}
\def\qone{q_1} \def\qtwo{q_2} \def\pone{p_1} \def\ptwo{p_2}
\def\qpm{q^\pm} \def\ppm{p^\pm}
\def\alplus{\alpha^+} \def\alminus{\alpha^-}
\def\alpm{\alpha^\pm}
\def\pplus{p^+} \def\pminus{p^-}

\def\alplusi{\alplus_{-1}}
\def\alminusi{\alminus_{-1}}
\def\vecali{\vec{\alpha}^f_{-1}}

\def\bminusi{b_{-1}}
\def\Aplus{A_+}  \def\Aminus{A_-}
\def\Kplus{K^+}  \def\Kminus{K^-}
\def\Nhatdlg{\hat{N}^{dLg}}
\def\invNhat{\hat{N}^{-1}_{dLg}}
\def\Nhatf{\hat{N}^f}
\def\np    { Nucl. Phys. }
\def\pr    { Phys. Rev. }
\def\pl    { Phys. Lett. }
\def\cmp   { Commun. Math. Phys. }
\def\jmp   { J. Math. Phys. }
\def\papertitlepage{\baselineskip 3.5ex \thispagestyle{empty}}
\def\Title#1{\vspace{2.5cm}\begin{center}
 {\Large\bf #1} \end{center}
\vspace{2cm}}
\def\Authors#1{\begin{center} {\it #1} \end{center}}
\def\Abstract{\vspace{2.7cm}\begin{center} {\large\bf Abstract}
           \end{center} \parbigskip}
\def\Komabanumber#1#2{\hfill \begin{minipage}{4cm} UT-Komaba #1
              \parn #2 \end{minipage}}

\renewenvironment{thebibliography}{\pagebreak[3]\par\vspace{0.6em}
\begin{flushleft}{\large \bf References}\end{flushleft}
\vspace{-1.0em}

\begin{enumerate}\if@twocolumn\baselineskip=0.6em\itemsep -0.2em
\else\itemsep -0.2em\fi\labelsep 0.1em}{\end{enumerate}}
\begin{document}
\papertitlepage
\Komabanumber{93-3}{March 1993}
\Title{Exact Operator Quantization of a Model of \\
\vskip 1.5ex Two-Dimensional Dilaton Gravity}
\Authors{{\sc\  S.~Hirano,\ \  Y.~Kazama\footnote[2]{e-mail address:\quad
kazama@tkyvax.phys.s.u-tokyo.ac.jp}\ \
 and\ \ Y.~Satoh} \\
 \vskip 3ex
 Institute of Physics, University of Tokyo, \\
 Komaba, Tokyo 153 Japan \\
  }
\Abstract
Exact operator quantization is perfomed of a model of two-dimensional
dilaton gravity in Lorentzian spacetime,  classically equivalent
to the one proposed by Callan, Giddings, Harvey and Strominger,
in the special case with 24 massless matter scalars.
This is accomplished by developing a
non-linear and non-local quantum canonical transformation of basic
interacting fields into a set of free fields, rigorously taking into
account the spatially closed
 boundary condition.  The quantized
 model enjoys conformal invariance and the entire set of physical
states and operators are obtained in the BRST formalism.  In addition,
  a rather detailed discussion of the nature of the basic issues for
exact treatment of models of quantum gravity is provided.
\newpage
\baselineskip=0.6cm
\section{Introduction} \sectionnumbering
The sentiment that we are on the verge of unveiling
(at least a part of) the long standing mystery of quantum gravity may
well turn out be too innocent, but the developments in the last few years
in low-dimensional quantum gravity appear to contain encouraging
evidences to make us feel tempted for such an optimism.  Powerful
matrix model techinique \cite{GM} opened up an
avenue for exact non-perturbative analysis and the target space
interpretation of the
 gauged Wess-Zunimo-Witten models\cite{w1} revealed a way in which
such an interesting and important physics as that of a black hole can be
unravelled in string theory.  \parsmallskip
 Encouraging signs are not confined in the realm of string theory.
Notably after the work of Callan, Giddings, Harvey and
Strominger (CGHS)\cite{CG}, it has been recognized that a class of field
theoretical models of quantum gravity in two dimensions,
characteristically containing a dilaton field, provide an excellent
 testing ground for a variety of fundamental issues in
 quantum gravity.
Semi-classical anaysis, which partially incorporates the back reaction
of the metric field, has been perfomed by many authors and demonstrated
the usefulness of such models \cite{CG}\nocite{RS1}\nocite{HW2}
\nocite{BD}\nocite{RS2}\nocite{STh}\nocite{BG}--\cite{BG}.
In particular, the phenomenon of evaporation of a black hole by
the emission of the Hawking radiation has been vigorously pursued with
fair amount of success.  \parsmallskip
However, semi-classical analysis of course
 has its limitations.  The type of analysis performed is valid only
 in the limit of large black hole mass and in the presence of a large
number of massless matter fields, and the approximation breaks down
 for the most interesting phase which determines the ultimate fate of
 a black hole.  Also in such a framework
 the difficult yet all important problem of the interpretation of
the wave function, including the question of unitarity and loss
 of quantum coherence, cannot properly be addressed. It is therefore
 clear that a more powerful treatment, with full-fledged quantization
 of the gravitational degrees of freedom, is desired.
\parsmallskip
There have been a number of attempts to quantize the CGHS model beyond
the semi-classical approximation, however with moderate success
\cite{AL}\nocite{ST}--\cite{HA3}.
One of the major difficulties is the choice of the proper functional
measure which defines the quantum theory.  Many of the procedures
 so far proposed lead, after a complicated non-linear field
transformation, to a free field theory, but the change of the
measure due to such a tranformation is to say the
least treated in an unclear manner.  One may take an attitude
 to regard the resultant free theory as defining a quantum model,
 but then the
quantum mechanical relations between these free fields and the original
interacting fields, for which we must make physical interpretations,
  will remain concealed.  A slightly different choice of the measure
 which permits more rigorous treatment has also been
proposed \cite{HA1}\nocite{HA2}--\cite{HA3}, but in this case there is
 a different problem; the quantum theory so obtained does not appear to
yield to analysis beyond semi-classical approximation.  \parsmallskip
The purpose of the present article is to give an improvement on this
situation by providing an interacting model for which the measure
chosen is clear  and at the same time exact operator quantization is
possible.
The model, to be defined precisely in the next section,
 can be regarded as  a special case of the CGHS model, namely the case
where the number of massless matter scalars is exactly 24, with a certain
 choice of the  measure.  To make the model well-defined
we shall impose spatially closed boundary conditions and stay in
Minkowski space throughout  in order to be able to discuss spacetime
 physics, including such important concepts as causality, locality, etc..
Exact operator quantization will then be accomplished by developing
 a non-linear and non-local canonical transformation, valid quantum
mechanically as well as classically, which maps the original
interacting fields to a set of free fields.  The model so quantized
 enjoys conformal invariance and by utilizing it we shall provide
a complete analysis of the physical states and operators in the
BRST framework.   In this article,  we will not be able to give a
 physical interpretation of our results  as there are still
many difficult problems to be overcome.  These problems, which we
believe must be faced in any serious attempt for exact
treatment, will be explained in detail.  In particular, the necessity of
a careful examination of the choice of the inner product for the space
of states is stressed.  Our emphasis throughout
 will be the rigor of the analysis, which we believe is particularly
 important for quantum gravity,  for various intuitions cultivated in our
experience with ordinary field theories must be attentively scrutinized.
 \parsmallskip
The rest of the paper is organized as follows.  In Sect.~2, we shall
define the model and study its classical properties.  A canonical
transformation is introduced and proved in Sect.~3 and with its
use the model is quantized in a rigorous manner. Conformal
properties of various fields will also be clarified.  Sect.~4 is devoted
 to the analysis of the physical states and operators of the theory
\'{a} la BRST.  Similarities to and differences from  the
case of non-critical closed string theory formulated in Euclidean space are
clarified.
 In Sect.~5, a rather detailed discussion of the nature of the remaining
problems
will be given.  Finally two appendices are
provided to supplement the technical details omitted in the text.
\section{The Model and its Classical Properties }
 \sectionnumbering
\subsection{ The Model }
The model we shall study in this article is classically identical to the
 one introduced by CGHS\cite{CG}.  Its action is given by
\eqabegin
    S &=& {1 \over \ff} \int d^2 \xi \sqrt{-g}\left\{\e^{-2{\phi}}\left[
    4g^{\alpha\beta} \partial_\alpha\phi\partial_\beta\phi
       +R_g  -4\lambda^2  \right]\
      + \sum_{i=1}^N \half g^{\alpha\beta}\partial_
         \alpha f_i\partial_\beta f_i\right\} \comma
   \label{eqn:cghs}
\eqaend
where $\phi$ is the dilaton field and  $f_i\ (i=1,\ldots, N) $ are
N massless scalar fields representing matter degrees of freedom.
 We shall stay throughout in Minkowski space and use the metric
convention
 such that for flat space $\eta_{\alpha\beta} = diag(1,-1)$.  In order
to define all the quantities unambiguously and to be able to perform
integrations by parts which occur in various places, we shall take
our universe to be spatially closed with period $2\pi L$.  For this
purpose, it is convenient to introduce the rescaled coordinates
\eqabegin
 x^\mu &= & (t,\sigma) = \xi^\mu / L \comma
\eqaend
and we require that all the fields appearing in the action be
periodic in $\sigma$, \ie,
\eqabegin
 F(t,\sigma+2\pi) &=& F(t, \sigma) \period
\eqaend
When the action is rewritten in terms of $x^\mu$, it retains its
form except with the replacement $\lambda \rightarrow \mu
 \equiv \lambda L$.  From now on, we will deal with such a
\lq\lq dimensionless" form and when necessary recover
the correct dimensions by appropriately multiplying by the factors
of $1/L$.  \parsmallskip
A choice of a classical action of course does not fix
a quantum theory.  We must specify the canonical variables and the form
of the measure to be used for the functional integration.
 Except for the requirement of invariance under general coordinate
transformations, there is no absolute maxim to be imposed by
general principle and hence the choice is far from unique.  In fact
a number of choices for the measure have been proposed and analyzed with
varying degree of rigor and  naturalness \cite{AL}\nocite{ST}\--\cite{HA3}.
\parsmallskip
 One attractive line of thought is to look for
 a choice such that the quantum thoery so obtained consistently retains
the on-shell conformal invariance present in the classical theory in
the conformal gauge.  The work of Bilal and Callan \cite{BC}, for
instance, represents such an attempt, although their procedure was
 somewhat indirect and not without obscurity.
  Recently a more transparent way of deriving
 a quantum theory with conformal invariance, which leads to
essentially the same model as that of Bilal and Callan,
 was proposed by
Hamada and Tuchiya \cite{HA3} (see in particular the appendix).
Their starting point is the action proposed by Russo and Tseytlin (RT)
 \cite{RT}\cite{OS}, which is classically equivalent to the CGHS
 action (\ref{eqn:cghs}).  It is obtained from (\ref{eqn:cghs})
 by the following transformation of fields:
\eqabegin
   \Phi &\equiv& \e^{-2\phi}\comma \\
    h_{\alpha\beta} &\equiv& \e^{2\omega}g_{\alpha\beta} \comma
\eqaend
where
\eqabegin
\omega &=& \frac12\left(\ln\Phi - \Phi\right) \period
\eqaend
Then the action becomes
\eqabegin
    S &=& \frac1{\ff}\int d^2x\sqrt{-h}\left[
    h^{\alpha\beta} \del_\alpha\Phi \del_\beta\Phi
      +R_h\Phi  -4\mu^2\e^\Phi + \sum_i \frac12 h^{\alpha\beta}\partial_
         \alpha f_i\partial_\beta f_i\right] \comma
   \label{eqn:rt}
\eqaend
where the curvature scalar $R_h$  refers to the conformally transformed
  metric $h_{\alpha\beta}$.  The authors of ref.\cite{HA3} take
the functional
measures for the fields $h_{\alpha\beta}$, $\Phi$ and $\vec{f}$ to be
those defined by the following norms:
\eqabegin
 \parallel \delta h \parallel^2 &=& \int d^2x \sqrt{-h}\,
   h^{\alpha\beta}h^{\gamma\delta}\delta h_{\alpha\gamma}h_{\beta\delta}
 \comma\nn\\
 \parallel \delta\Phi \parallel^2 &=& \int d^2x \sqrt{-h}\,
  \delta\Phi \delta\Phi \comma \label{eqn:measure} \\
 \parallel \delta f_i \parallel^2 &=& \int d^2x \sqrt{-h}\,
  \delta f_i \delta f_i  \qquad (i=1,\ldots, N) \period\nn
\eqaend
Then they separate $h_{\alpha\beta}$ into the Weyl factor
$\rho$ and the background metric $\ghat_{\alpha\beta}$
 as $h_{\alpha\beta} =\e^{2\rho}\ghat_{\alpha\beta}$ and rewrite
the measure into the one with respect to $\ghat$ \'{a} la David and Distler
 and Kawai \cite{DA}\cite{DK}.  The resulting action takes the
form
\eqabegin
S &=& {1\over \ff}\int d^2x \sqrt{-\ghat}\, \left[\, \ghat^{\alpha\beta}
 \del_\alpha\Phi\del_\beta\Phi + 2\ghat^{\alpha\beta}\del_\alpha\Phi
 \del_\beta \rho + R_{\ghat}\Phi -4\mu^2 \e^{\Phi + 2\rho}
 \right. \nn\\
 & & + \left. \kappa \left( \ghat^{\alpha\beta}\del_\alpha\rho
 \del_\beta\rho + R_{\ghat}\rho\right)
  +\half\ghat^{\alpha\beta}\del_\alpha\vec{f} \cdot \del_\beta\vec{f}
\, \right] + S^{gh}(\ghat, b,c) \label{eqn:ht} \comma
\eqaend
where $\kappa = (N-24)/12$ and $S^{gh}$ is the usual action
 for the reparametrization ghosts $b$ and $c$.  For $\kappa
 \ne 0$, they further perform a simple redefinition of fields ( with
 a unit Jacobian ) and show that the result is essentially identical
to the effective action obtained by Bilal and Callan.  \parsmallskip
Because of the transparency of its derivation, we shall take this
action (\ref{eqn:ht}) seriously and analyze in this article the
special case with $N=24$ in detail.  Although the action still contains
an exponential interaction, we shall see that it can be canonically
mapped to a free theory and hence can be solved exactly. \parsmallskip
In this sense the model is certainly mathematically consistent.
we need, however,  to mention two subtleties to be kept in mind.
 The first is the  question of the range of $\Phi$ in the functional
integration.  Classically $\Phi$ is a non-negative quantity and as was
emphasized in \cite{HA3} there is no symmetry principle which allows
one to naturally extend this range into the negative region.
Ignoring this restriction may or may not be
a serious problem and the answer should await a detailed analysis.
The second concerns the physical interpretation of the model.  Namely,
 the procedure  outlined above can be interpreted in two ways.
One point of view
 is to take the classical RT action seriously and regard
$h_{\alpha\beta}$
 as the genuine metric of the spacetime.  Then the choice of the
measure (\ref{eqn:measure}) seems  natural.  An alternative
 view is to regard the procedure as quantizing the original
 CGHS action with a definite but somewhat unusual measure.
In this standpoint, one
 continues to interpret $g_{\alpha\beta}$ as the metric.  Again the
justification of one or the other can only be decided after a detailed
 study of the model. \parsmallskip
These subtleties will have to be watched but we believe that the model
has a big  advantage in that one knows the precise setting and that
it is still solvable.
\subsection{Classical Properties }
Since the ghost part of the action can be handled in the usual way, we
first look at the remainder, which we shall call the classical action
 $S^{cl}$ .
Setting the reference metric $\ghat_{\alpha\beta}$ to be the
flat metric $\eta_{\alpha\beta}$ and $N$ to be 24, the classical action
 becomes
\eqabegin
 S^{cl} &=& {1\over \ff} \int d^2x \left( \del_\alpha\Phi\del^\alpha\Phi
 + 2\del_\alpha\Phi\del^\alpha\rho -4\mu^2\e^{\Phi + 2\rho}
 + \half \del_\alpha \vec{f} \cdot \del^\alpha \vec{f} \right) \period
 \label{eqn:cl}
\eqaend
Hereafter we shall set $\ff=1$ for simplicity.
Variation with respect to $\Phi$, $\rho$ and $\vec{f}$ gives
the following equations of motion:
\eqabegin
 2\delplus\delminus \Phi + 2\delplus\delminus \rho
  + \mu^2 \e^{\Phi + 2\rho} &=& 0 \comma
       \label{eqn:Phieq} \\
 \delplus\delminus \Phi + \mu^2 \e^{\Phi + 2\rho} &=& 0 \comma
       \label{eqn:rhoeq} \\
 \delplus\delminus \vec{f} &=& 0 \period \label{eqn:feq}
\eqaend
Here and hereafter the light cone coordinates are defined by
\eqabegin
 x^\pm = x^0\pm x^1 = t\pm \sigma \comma \qquad
 \del_\pm = \half(\del_t \pm \del_\sigma)\comma \qquad
 \Box = 4\delplus\delminus \period
\eqaend
General solutions for these equations of motion can easily be obtained:
First, each matter scalar $f_i$ is trivially a free field.  Next,
 by eliminating the exponential term from Eqs.(\ref{eqn:Phieq})
and (\ref{eqn:rhoeq}),  one finds that $\Phi + 2\rho\equiv \psi
 = \psi^+ + \psi^-$ is a free field.
Next put this back into Eq.(\ref{eqn:rhoeq}) and define the functions
 $A(\xplus)$ and $B(\xminus)$ by
\eqabegin
 \delplus A(\xplus)&=& \mu \e^{\psi^+(\xplus)} \comma \label{eqn:Aeq}\\
 \delminus B(\xminus) &=& \mu\e^{\psi^-(\xminus)}\period \label{eqn:Beq}
 \eqaend
Then it is easy to see that $\Phi +AB \equiv \chi$ is again a free field.
Thus we can write
\eqabegin
 \Phi &=& \chi -AB \comma\label{eqn:Phisol} \\
 \rho &=& \half ( \psi -\Phi )\period \label{eqn:rhosol}
\eqaend
Since $\psi$ and $\chi$ satisfy periodic boundary conditions, we can
expand them into Fourier modes as follows:
\eqabegin
 \psi^\pm &=& \half Q_\psi + {P_\psi \over 4\pi}x^\pm
  + {i\over \sqrt{4\pi}}\sum_{n\ne 0}\alpha^\pm _n \e^{-inx^\pm}
 \comma \label{eqn:psimode} \\
 \chi^\pm &=& \half Q_\chi + {P_\chi \over 4\pi}x^\pm
  + {i\over \sqrt{4\pi}}\sum_{n\ne 0}\beta^\pm _n \e^{-inx^\pm}\period
  \label{eqn:chimode}
\eqaend
\parmedskip
We now solve the equtations for $A(\xplus)$ and $B(\xminus)$.
 Although the product $AB$ must be a periodic function,
$A(\xplus)$ and $B(\xminus)$ separately need not be
periodic.  In fact, since the left- and right-moving components
 $\psi^\pm$ each undergoes a constant shift, the correct boundary
conditions for $A(\xplus)$ and $B(\xminus)$ are of the form
\eqabegin
 A(\xplus +2\pi) &=& \alpha A(\xplus) \comma\\
 B(\xminus -2\pi) &=& {1\over \alpha} B(\xminus) \period
\eqaend
{}From the mode expansion of $\psi$, the  constant $\alpha$ is
easily seen to be related to the momentum zero mode $P_\psi$ by
\eqabegin
 \alpha &=& \e^{P_\psi /2} \period
\eqaend
  Now we observe that the equations as well as
the boundary conditions above for $A(\xplus)$ and $B(\xminus)$ are
identical in form
to those which appeared in the operator analysis of the Liouville theory
 \cite{OW}\cite{KN}.  Suppressing
 the dependence on $t$, the solutions can be expressed as
\eqabegin
 A(\sigma) &=& \mu C(\alpha) \int_0^{2\pi} d\sigma'
    E_\alpha(\sigma-\sigma') \e^{\psi_+(\sigma')} \comma\\
 B(\sigma) &=& \mu C(\alpha)\int_0^{2\pi} d\sigma''
    E_{1/\alpha}(\sigma-\sigma'') \e^{\psi_-(\sigma'')} \comma
\eqaend
where $C(\alpha) =1/\left(\sqrt{\alpha}-\sqrt{\alpha}^{-1}\right)$
 and the functions $E_\alpha(\sigma)$ and $E_{1/\alpha}(\sigma)$
 are defined by
\eqabegin
  E_\alpha(\sigma) &\equiv&
      \exp\left(\half\ln\alpha\epsilon(\sigma)\right) \comma\\
  E_{1/\alpha}(\sigma) &\equiv&
      \exp\left(-\half\ln\alpha\epsilon(\sigma)\right) \period
\eqaend
$\epsilon(\sigma)$ is a stair-step function with the property
\eqabegin
 \epsilon(\sigma +2\pi) &=& 2 +\epsilon(\sigma) \comma
\eqaend
and it coincides with the usual $\epsilon$-function in the interval
 $[-2\pi, 2\pi]$.  It is useful to note  that the derivatives of
 $E_\alpha(\sigma)$ and $E_{1/\alpha}(\sigma)$ are proportional
 to the  periodic $\delta$ function, {\it viz.},
\eqabegin
 \delsigma E_\alpha(\sigma-\sigma') &=&
     {1\over C(\alpha)}\delta(\sigma-\sigma') \comma\\
 \delsigma E_{1/\alpha}(\sigma -\sigma') &=&
     -{1\over C(\alpha)}\delta(\sigma-\sigma') \period
\eqaend
\parmedskip
In addition to the equations of motion discussed above, there are
constraint equations which follow from general covariance, namely
those expressing the vanishing of the energy-momentum tensor
$T_{\alpha\beta}$, which is obtained by varying the
action (\ref{eqn:ht}) with respect to the reference metric $\ghat$.
 Classical parts of $T_{\alpha\beta}$, after setting $\ghat_{\alpha\beta}
 = \eta_{\alpha\beta}$, are given in the light-cone coordinates by
\eqabegin
 T_{\pm\pm}&=& (\partial_\pm \Phi)^2
  -\partial_\pm^2 \Phi +2\partial_\pm \rho \partial_\pm \Phi +\frac12
        (\partial_\pm \vec{f})^2 \comma\label{eqn:fEMT}\\
\
 T_{+-}&=& \partial_+ \partial_- \Phi +\mu^2 e^{\Phi+2\rho}
     \period
\eqaend
{}From the previous equation of motion for $\rho$ (Eq.(\ref{eqn:rhoeq}))
 $T_{+-}$ is seen to vanish, showing the on-shell conformal invariance
 at the classical level.  As for $T_{\pm\pm}$, use of
Eqs.(\ref{eqn:Phisol}) and (\ref{eqn:rhosol}) readily yields
\eqabegin
 T_{\pm\pm} &=& \del_\pm \chi \del_\pm \psi -\del_\pm ^2\chi
   + \half (\del_\pm \vec{f})^2 \period
 \eqaend
Defining $\tilde{\phi}_1$ and $\tilde{\phi}_2$ by
\eqabegin
 \psi &=& \overrttwo (\tilde{\phi}_1 +\tilde{\phi}_2)
   \comma \label{eqn:phidefone}\\
 \chi &=& \overrttwo (\tilde{\phi}_1 -\tilde{\phi}_2)
\comma \label{eqn:phideftwo}
\eqaend
it can be diagonalized as
\eqabegin
  T_{\pm\pm}&=& \half( \del_\pm \tilde{\phi}_1 )^2
-\overrttwo\del_\pm^2 \tilde{\phi}_1
  +\half( \del_\pm \tilde{\phi}_2 )^2 +\overrttwo\del_\pm^2
 \tilde{\phi}_2  +\half (\del_\pm \vec{f})^2 \period \label{eqn:ffem}
\eqaend
Thus the vanishing of $T_{\pm\pm}$ simply relates the chiral components
 of the three types
 of free fields $\chi$, $\psi$ and $\vec{f}$.  In order to enforce the
proper boundary conditions, we find it advantageous to take
the functions $A(\xplus)$, $B(\xminus)$ and $\chi$ as arbitrary
 and compute the corresponding $\psi$ and $\vec{f}$. \par
$A(\xplus)$ and $B(\xminus)$ with the proper boundary conditions can
be written as
\eqabegin
 A(\xplus) &=& \mu \e^{(P_\psi/4\pi) \xplus} a(\xplus)\comma\\
 B(\xminus) &=& \mu \e^{(P_\psi/4\pi) \xminus}b(\xminus)\comma
\eqaend
where $a(\xplus)$ and $b(\xminus)$ are arbitrary periodic functions.
Then by simple calculations, $\psi^\pm(x^\pm)$ and
$(\del_\pm \vec{f})^2$ can be expressed as
\eqabegin
 \psi_+(\xplus)
 &=& {P_\psi \over 4\pi}\xplus + \ln \left( {P_\psi \over 4\pi}
a(\xplus)+\delplus a(\xplus) \right) \comma\\
 \psi_-(\xminus) &=& {P_\psi \over 4\pi}\xminus +
\ln \left( {P_\psi \over 4\pi}b(\xminus)+\delplus b(\xminus) \right)
 \comma\\
 \half (\delplus f)^2
 &=& -\left({P_\psi \over 4\pi}+{{P_\psi \over 4\pi}\delplus a
 + \delplus^2 a \over {P_\psi \over 4\pi} a + \delplus a}\right)
 \delplus\chi +\delplus^2 \chi \comma\\
 \half (\delminus f)^2
 &=& -\left({P_\psi \over 4\pi}+{{P_\psi \over 4\pi}\delminus b
 + \delminus^2 b \over {P_\psi \over 4\pi} a + \delminus b}\right)
 \delminus\chi +\delminus^2 \chi \period
\eqaend
In terms of these quantities, the original metric $g_{\alpha\beta}$
is given by
\eqabegin
 g_{\alpha\beta} &=& \e^{\psi}\left(\chi -AB\right)^{-1}
 \eta_{\alpha\beta}\nn\\
 &=& \e^{\psi}\left(\chi-\mu^2 \e^{(P_\psi/ 2\pi) t}a(\xplus)b(\xminus)
\right)^{-1}\eta_{\alpha\beta} \period \label{eqn:orgmet}
\eqaend
This is {\it the finite-universe version} of the general solution
obtained  by CGHS\cite{CG}. \parmedskip
Let us give a simple example which describes a black hole metric in the
limit that the size of the universe $L$ tends to infinity.  It is given
 by the choice
\eqabegin
 \chi &=& c = \mbox{constant}, \qquad \del_\pm \vec{f} = 0 \comma\\
 a(\xplus) &=& \sin\xplus,\qquad b(\xminus) = \sin\xminus \period
\eqaend
The corresponding $\psi^\pm$ are given by
\eqabegin
 \psi_\pm &=& {P_\psi \over 4\pi}x^\pm + \ln \left({P_\psi \over 4\pi}
 \sin x^\pm + \cos x^\pm \right) \period
\eqaend
Notice that in the limit $L\rightarrow \infty$,\  $\psi^\pm$ vanish.
Recalling that $\mu = \lambda L$, the metric
becomes, in this limit
\eqabegin
 \lim_{L\rightarrow \infty} g_{\alpha\beta} &=&
 \lim_{L\rightarrow \infty}\left[ \e^{\psi}
 \left(c -\mu^2 \e^{(P_\psi/ 2\pi) t}\sin \xplus
 \sin \xminus\right)^{-1}\eta_{\alpha\beta} \right] \nn\\
  &=&  \left(c -\lambda^2 \xi^+ \xi^-\right)^{-1}\eta_{\alpha\beta}
 \comma
\eqaend
which describes a black hole.  We must note here that in fact infinitely
 many other configurations lead to the same $L\rightarrow\infty$ limit.
Specifically, any choices of $a(\xplus)$ and $b(\xminus)$ for which
$L\, a(\xplus)$ and $L\, b(\xminus)$ tend, respectively, to $\xi_+$ and
$\xi_-$ are  indistinguishable in the limit of a large universe.
 We will have more to say on this point in the final section.
%
\section{ Quantization of the Model } \sectionnumbering
\subsection{Canonical Transformation and Operator Quantization }
In the previous section, we found that the general solution to
the classical equations of motion can be described by three types of
free fields, $\psi$, $\chi$ and $\vec{f}$.  Quantization of the matter
sector is trivial since $f_i$'s are canonical right from the beginning.
 For the dilaton-Liouville sector, however, it is not yet obvious what
combinations of $\psi$ and $\chi$ are to be quatnized as {\it canonical}
free fields for the following reasons:  First, any function of free
 fields is again a free field and depending on the choice an additional
 functional Jacobian may arise.  Moreover, the form of the
energy-momentum tensor $T_{\pm\pm}$ alone is in general not sufficient
 to settle the question.  After all, the fact that they take \lq\lq
free-field form", as in (\ref{eqn:fEMT}), does not even guarantee that
 the the fields appearing in them are free fields.  \parsmallskip
 Nevertheless, the  expression obtained in (\ref{eqn:ffem}) is
suggestive.  It appears to imply that $\tilde{\phi}_1$ and $\tilde{\phi}
_2$ should be regarded as canonical free fields and that, while the
former being a normal field, the latter should be treated as a \lq\lq
negaive metric" field.  ( This should not
be taken as dictating the way the inner product should be defined.  It
is a separate problem to be discussed in the final section. )
  In this subsection, we shall prove rigorously that this expectation
is indeed correct. Namely, it will be shown that the transformation
 of fields $(\Phi, \rho) \rightarrow (\tilde{\phi}_1,\tilde{\phi}_2)$ is
 canonical quantum mechanically as well as classically. \parsmallskip
Let us begin with classical analysis.  First, for later convenicence, we
shall rescale the fields  $\tilde{\phi}_1$ and $\tilde{\phi}_2$ to
define canonically normalized fields $\phione$ and $\phitwo$  and
expand them into Fourier modes  as follows:
\eqabegin
 \phi_i &\equiv& \sqrt{4\pi}\, \tilde{\phi}_i \comma\\
 \phi_i(\xplus,\xminus) &=& \phi_i^+(\xplus) + \phi_i^-(\xminus)
 \comma\\
 \phi_i^\pm(x^\pm) &=& {q^i \over 2} + p^i x^\pm
    + i\sum_{n\ne 0} {1\over n} \alpha_n^{(i,\pm)}
    \e^{-inx^\pm} \period \label{eqn:Frphi}
\eqaend
We take the basic Poisson brackets to be
\eqabegin
  i\left\{ \alpha_m^{(1,\pm)}, \alpha_n^{(1,\pm)} \right\} &=&
  m\delta_{m+n,0} \comma \\
\left\{q^1, p^1 \right\} &=& 1 \comma\\
   i\left\{ \alpha_m^{(2,\pm)}, \alpha_n^{(2,\pm)} \right\}
 &=& - m\delta_{m+n,0} \comma\\
\left\{q^2, p^2 \right\} &=& -1 \comma \\
 \mbox{Rest} &=& 0 \period
\eqaend
After some calculations, this leads to
\eqabegin
 \left\{\phi_1^+(\xplus), \phi_1^+(\yplus) \right\}
  &=& \half(\xplus -\yplus) -\pi \epsilon(\xplus
     -\yplus) \comma\\
 \left\{\phi_1^-(\xminus), \phi_1^-(\yminus) \right\}
  &=& \half(\xminus -\yminus) -\pi \epsilon(\xminus
     -\yminus) \comma\\
 \left\{\phi_1^+(\xplus), \phi_1^-(\yminus) \right\}
  &=& \half(\yminus -\xplus) \comma\\
 \left\{\phi_1^-(\xminus), \phi_1^+(\yplus) \right\}
  &=& \half(\yplus -\xminus) \comma\\
 \left\{ \phi_1(x), \phi_1(y) \right\}
   &=& -\pi \left( \epsilon(\xplus-\yplus)
      -\epsilon(\xminus -\yminus)\right) \comma
\eqaend
and similar expressions with all the signs reversed for $\phitwo$.
 ( The $\epsilon$ function here is the stair-step function defined in
the previous section. ) \parmedskip
In terms of $\phi_i$, the fields $\psi$ and $\chi$ now take the form
\eqabegin
 \psi &=& {1\over \sqrt{8\pi}}(\phione +\phitwo )\comma\\
 \chi &=& {1\over \sqrt{8\pi}}(\phione -\phitwo )\comma
\eqaend
and their Fourier modes, defined in Eqs.(\ref{eqn:psimode}),
(\ref{eqn:chimode}),
are easily seen to satisfy the Poisson bracket relations
\eqabegin
i\left\{ \alpha_m^\pm , \beta_n^\pm \right\}
  &=& i\left\{ \beta_m^\pm , \alpha_n^\pm \right\}
  = m\delta_{m+n, 0}  \comma\\
 \left\{Q_\psi, P_\chi\right\} &=& 1 \comma\\
 \left\{Q_\chi, P_\psi\right\} &=& 1 \comma\\
 \mbox{Rest} &=& 0 \period
\eqaend
Then with the use of previous formulae for $\phi_i$, the following
 brackets are readily obtained:
\eqabegin
 \left\{ \chi(x), \psi^+(y)\right\}
&=& {1\over 8\pi}(\xplus -\xminus) -\quarter \epsilon(\xplus
     -\yplus) \comma \label{eqn:chipsi}\\
 \left\{ \chi(x), \psi^-(y)\right\}
&=& -{1\over 8\pi}(\xplus -\xminus) -\quarter \epsilon(\xminus
     -\yminus) \comma\\
 \left\{ \psi(x), \psi(y)\right\} &=& \left\{ \chi(x), \chi(y)\right\}
    =0 \comma\\
\left\{ \psi(x), \chi(y) \right\} &=& \left\{ \chi(x), \psi(y) \right\}
  = -{1\over 4}\left(\epsilon(\xplus -\yplus)
          +\epsilon(\xminus-\yminus)\right) \period \label{eqn:psichi}
\eqaend
\par
We are now ready to prove the canonical nature of the transformation.
The Poisson brackets we wish to reproduce {\it at equal time} are
\eqabegin
  \left\{ \Phi(x),\Pi_\Phi(y) \right\}_{ET}
    &=& \delta(\sigma_x -\sigma_y)\comma \nn\\
\left\{ \rho(x), \Pi_\rho(y) \right\}_{ET}
    &=& \delta(\sigma_x -\sigma_y) \comma\label{eqn:canPB}\\
 \mbox{Rest} &=& 0 \comma\nn
\eqaend
where the momentum fields are given by
\eqabegin
 \Pi_\Phi &=& 2 (\Phidot + \rhodot) \comma\\
 \Pi_\rho &=& 2  \Phidot \period
\eqaend
Thanks to the fact that the modes of $\psi$ have vanishing Poisson
brackets  with each other,
the only non-trivial brackets to be computed are
 $ \left\{ \chi(x), AB(y) \right\}  $ and the time derivatives thereof
 at equal time.  Calculations are somewhat tedious and very similar
to the ones needed for the self-interacting Liouville theory
\cite{KN}.
 In particular one needs to be careful about the presence of the
momentum zero mode in the functions $E_\alpha(\sigma)$ and
 $E_{1/\alpha}(\sigma)$ inside $AB$.  Useful formulae are listed in
the appendixA.  \par
With the help of these formulae, it is straightforward to get the
following brackets:
\eqabegin
 \left\{ \Phi(x), \Phi(y) \right\}_{ET} &=&
  - \left\{ \chi(x), AB(y) \right\}_{ET} -
    \left\{ AB(x), \chi(y) \right\}_{ET} = 0 \comma\\
 \left\{ \Phidot(x), \Phi(y) \right\}_{ET} &=& 0 \comma\\
 \left\{ \Phidot(x), \Phidot(y) \right\}_{ET} &=& 0 \comma\\
 \left\{ \rho(x), \rho(y) \right\}_{ET} &=&
      -\frac1{4} \Biggl[ \left\{ \chi(x), \psi(y) \right\}_{ET}
      +\left\{ \psi(x), \chi(y) \right\}_{ET} \Biggr] = 0  \comma\\
 \left\{ \rhodot(x), \rho(y) \right\}_{ET} &=&
       \half \delta( \sigma_x - \sigma_y ) \comma\\
 \left\{ \rhodot(x), \rhodot(y) \right\}_{ET} &=& 0 \comma\\
 \left\{ \Phi(x), \rho(y) \right\}_{ET} &=& \half \left\{ \chi(x),
      \psi(y) \right\}_{ET} = 0 \comma\\
 \left\{ \Phidot(x), \rho(y) \right\}_{ET} &=& -\half
 \delta( \sigma_x - \sigma_y ) \comma\\
 \left\{ \Phi(x), \rhodot(y) \right\}_{ET} &=&  \half
\delta( \sigma_x - \sigma_y ) \comma\\
 \left\{ \Phidot(x), \rhodot(y) \right\}_{ET} &=& 0 \period
\eqaend
{}From these bracket relations, it is evident that our transformation
correctly reproduces the canonical Poisson bracket relations
(\ref{eqn:canPB}).
\parsmallskip
Now we can quantize the theory by the replacement
\eqabegin
 \left\{\phi_i, \phi_j\right\}_{ET} &\longrightarrow &
  (-i)\left[\phi_i, \phi_j\right] \period
\eqaend
To prove the canonical nature of the transformations quantum
mechanically, we must define the composite operator $AB$.  Here
we have a situation far simpler than the corresponding case for
the Liouville theory \cite{OW}\cite{KN}: All the modes
 of $\psi$ commute with each other and $AB$ is well-defined without
the need of normal ordering.    Thus all the Poisson bracket
relations previously obtained can be directly converted into
quantum commutation relations, and the quantum canonicity follows
 immediately from the classical one.
\subsection{Conformal Properties}
%
As was shown in subsection 2.2, the off-diagonal part of the energy-
momentum tensor $T_{+-}$ vanishes due to the equations of motion and
the model has invariance under the left- and right- conformal
transformations.  Upon quantization the left-Virasoro
 generators for the dilaton-Liouville and the matter sectors
take the form
\eqabegin
 \Ldl_n &=& L_n^1 + L_n^2 \comma\label{eqn:Ldl} \\
 L_n^1 &=& \half \sum_m :\alpha^1_{n-m}\alpha^1_m:+iQn\alpha^1_n\comma
 \label{eqn:Lone}\\
 L_n^2 &=& -\half \sum_m :\alpha^2_{n-m}\alpha^2_m:-iQn\alpha^2_n\comma
 \label{eqn:Ltwo}\\
 \Lf_n &=&  \half \sum_m :\vecal^f_{n-m}\cdot\vecal^f_m: \comma
 \label{eqn:Lf}
\eqaend
where the background charge $Q$ is give by $Q = \sqrt{2\pi}$.
Paying attention to the negative metric nature of $\alpha^2_n$'s, we
 readily obtain
\eqabegin
\left[ L^1_m, L^1_n \right] &=& (m-n)L^1_{m+n}
   + {1+12 Q^2 \over 12}(m^3-m)\deltamn + Q^2m \deltamn \comma
  \label{eqn:virone}\\
\left[ L^2_m, L^2_n \right] &=& (m-n)L^2_{m+n}
   + {1-12 Q^2 \over 12}(m^3-m)\deltamn - Q^2m \deltamn \comma
 \label{eqn:virtwo}\\
 \left[\Ldl_m, \Ldl_n \right] &=& (m-n)\Ldl_{m+n}
  + {2\over 12}(m^3-m)\deltamn \comma \label{virdl}\\
 \left[\Lf_m, \Lf_n \right] &=& (m-n)\Lf_{m+n}
  + {N\over 12}(m^3-m)\deltamn \period \label{eqn:virf}
\eqaend
Thus, both $\Ldl_n$ and $\Lf_n$ satisfy the standard form of the
Virasoro algebra with the central charge 2 and N respectively and
together with the ghost contribution the total conformal anomaly
 vanishes for $N=24$.  \parsmallskip
Let us now discuss the conformal properties of the basic fields and
the operators involving their exponentials.  First, for the positive
metric field $\phione$, we easily find
\eqabegin
 \left[L^1_m, \phione(x)\right]&=&\e^{im\xplus}
 \left( \overi \delplus \phione + iQm \right) \period\label{eqn:confone}
\eqaend
As for the negaive metric field $\phitwo$, the sign of $L_m$
as well as those of the commutation relations are reversed, and the net
result is identical with the positive metric case.  Combining them
 we immediately get
\eqabegin
 \left[L_m, \psi(x)\right] &=& \e^{im\xplus}
 \left( \overi \delplus \psi +  i{Q\over \sqrt{2\pi}}m \right) \comma\\
 \left[L_m, \chi(x)\right] &=& \e^{im\xplus}
  \overi \delplus \chi  \period
\eqaend
Note that $\chi$ field transforms as a genuine primary field with
conformal dimension 0.   \parsmallskip
Next consider the simple exponential operator of the form $\e^{\lambda
\phi}$ where $\phi$ is either $\phione$ or $\phitwo$.  For the
\lq\lq cylinder" type coordinates under use for our spatially closed
Lorentzian universe, the proper normal ordering is the {\it symmetric}
 normal ordering defined by
\eqabegin
 :\e^{\lambda\phi(x)}: &=& \e^{\lambda q /2}
 \e^{p (\xplus+\xminus)}\e^{\lambda q /2} \nn\\
 & & \cdot \e^{\lambda \phi^+_c(\xplus)}\e^{\lambda\phi^+_a(\xplus)}
 \e^{\lambda \phi^-_c(\xminus)} \e^{\lambda\phi^-_a(\xminus)} \comma
\eqaend
where $\phi^\pm_a$ and $\phi^\pm_c$ refer respectively to the
annihiliation and creation part of the non-zero modes.
If we adopt the usual hermiticity assignment for the modes this operator
 is manifestly hermitian for real $\lambda$.  More importantly, it
becomes a primary field only with this normal ordering.
  As we cannot directly make use of the Euclidean operator product
 technique, the calculation of the commutator $\left[L_n, \e^{\lambda
 \phi}\right]$ is slightly tedious.  However the results are standard:
 For $\phione$ and $\phitwo$ we obtain
\eqabegin
 \left[ L^1_n, \e^{\lambda\phione(x)}\right]
 &=& \e^{in\xplus} \left( \overi \delplus
   + n \left(-{\lambda^2\over 2} + Q\lambda \right)\right)
 \e^{\lambda\phione(x)} \comma\label{eqn:cofexpone}\\
\mbox{dim}\ \e^{\lambda\phione} &=& -{\lambda^2\over 2} +
 Q\lambda \comma\label{eqn:dimexpone}\\
 \left[ L^2_n, \e^{\lambda\phitwo(x)}\right]
 &=& \e^{in\xplus} \left( \overi \delplus
   + n \left({\lambda^2\over 2} + Q\lambda \right)\right)
 \e^{\lambda\phitwo(x)} \comma\label{eqn:cofexptwo}\\
\mbox{dim}\ \e^{\lambda\phitwo} &=& {\lambda^2\over 2} +
 Q\lambda \period\label{eqn:dimexptwo}
\eqaend
This immediately implies that $\e^{\lambda\psi}$ and $\e^{\lambda\chi}$
 are primary fields with conformal dimensions
\eqabegin
 \mbox{dim}\ \e^{\lambda\psi} &=& {Q\over \sqrt{2\pi}}\lambda
  = \lambda \comma\\
 \mbox{dim}\ \e^{\lambda\chi} &=& 0 \qquad \mbox{(independent of
 $\lambda$) }\period
\eqaend
\parmedskip
%
Finally we need to clarify the conformal property of the composite
operator $AB(x)$.  As remarked in subsection 2.2, this operator consists
of modes of $\psi$ only and there is no necessity of normal ordering.
The calculation of the commutator with the Virasoro generator is rather
 tedious due to the non-local nature of the operator.
 However, if we formally adopt the symmetric normal ordering it becomes
almost identical to the corresponding  case encountered in the
interacting Liouville theory \cite{KN} and we can easily
transcribe their procedure.  The final result is
\eqabegin
 \left[ L^{dL}_n, AB(x)\right] &=& \e^{in\xplus}
 {1\over i} {\del \over \del\xplus} (AB(x))
  \comma\label{eqn:confAB}
\eqaend
showing that $AB$ is a primary field with dimension zero just like the
 field $\chi$.  It means that our fundamental field
$\Phi = \chi -AB$ as a whole behaves as a dimension zero primary field,
 a gratifying result.
%
\section{BRST Analysis of Physical States and Operators}
\sectionnumbering
Having quantized the model in a rigorous manner, we now construct the
physical states  and the operators of the theory in the BRST formalism.
  Because of the conformal symmetry, the anaysis will be quite similar
to the one for the string theory. In fact, technically, our model has
features which are hybrid of the critical and non-critical string
theories: It can be regarded as a critical bosonic string thoery since
 the total central charge is made up of the contributions from
 26 free bosons, with matter fields playing the role of the 24
transverse coordinates.   On the other hand, the presence of the
background charges for the remaining two coordinates leads to structures
reminiscent of the $c=1$ non-critical string theory.  \par
 Thus in the following, we shall be able to make use of the anaysis
performed on the non-critical string theory \cite{lz}\cite{bmp},
albeit with some modifications.
 These modifications are  due
({\it i}) to the fact that we stay in Minkowski space, ({\it ii}) to the
extra presence of the matter fields and ({\it iii}) to the special
structure of the background charge terms for the dilaton-Liouville
sector.  Our analysis is closely analogous to those performed for
non-interacting Liouville gravity \cite{Bl} and for a free field model
 of dilaton gravity \cite{Sakai} both defined in Euclidean space,
 but we believe it is useful to give some details and clarify the
difference between Minkowski and Euclidean formulations.  This will
 also serve to make this article sufficiently self-contained.
%
\subsection{Analysis of Physical States }
\subsubsection{Preliminary}
 In this section, we shall deal explicitly with the left-moving
sector only and $\phi_i$ will mean the chiral components with the
 following mode expansion and the commutation relations:
\eqabegin
 \phi_i (\xplus) &=& \half q_i + p_i\xplus + i\sum_{n\ne 0}
 {1\over n} \alpha_n^i \e^{-in\xplus} \\
 \delplus \phi_i &=& \sum_{n\in {\bf Z}} \alpha_n^i \e^{-in\xplus}
 \qquad ( \alpha^i_0 \equiv p^i )  \\
 \left[ q_1, p_1 \right] &=& i, \qquad
 \left[ q_2, p_2 \right] = -i \\
 \left[ q_1, p_2 \right] &=& \left[ q_2, p_2 \right]=0 , \\
 \left[ \alpha_m^1, \alpha_n^1 \right] &=&
 - \left[ \alpha_m^2, \alpha_n^2 \right] = m\deltamn , \qquad
 \left[ \alpha_m^1, \alpha_n^2 \right] = 0  \period
\eqaend
Since the Virasoro generators $\Ldl_n$ and $\Lf_n$ given in
Eqs.(\ref{eqn:Ldl}) $\sim$ (\ref{eqn:Lf}) both satisfy the standard
form of the Virasoro algebra with central charges 2 and 24 respectively,
 the nilpotent BRST operator, which we shall denote by $d$, is
immediately obtained as
\eqabegin
 d &=& \sum c_{-n} (\Ldl_n+\Lf_n) -\half
\sum :(m-n)c_{-m}c_{-n}b_{m+n} : \comma \label{eqn:BRSop}
\eqaend
where $sl(2)$ invariant normal-ordering for the ghosts is assumed.
The physical ghost vacuum is defined as usual by
 $ \downvacket = c_1 \mid 0 >_{inv} $.
\parsmallskip
Hereafter we shall follow closely the work of Bouwknegt, McCarthy
 and Pilch (BMP) \cite{bmp}.  Let us recall their general
strategy.  First the BRST operator $d$ is decomposed
 with respect to the ghost zero mode in the form
$d = c_0 L_0 -Mb_0 + \dhat$, where $L_0$ is the total Virasoro operator
 including the ghosts.
{}From the well-known relation $L_0 = \left\{b_0, d\right\}$
one can deduce by a standard argument that the non-trivial
$d$-cohomology must be in the sector satisfying
\eqabegin
 L_0 \psi &=& 0 \period
\eqaend
On the subspace $\calF_0$ defined by
\eqabegin
 \calF_0 &=& \left\{ \psi \mid L_0\psi =0, \quad b_0\psi =0 \right\}
 \comma
\eqaend
$\dhat$ becomes nilpotent and the $d$-cohomology (absolute cohomology)
is reduced to the $\dhat$-cohomology (relative cohomology). \parsmallskip
The next step is to further decompose the $\dhat$ operator with respect
to a grading called the {\it degree} to be assigned to the mode
operators.  For this purpose, define the following light-cone-like
 combinations of modes for the dilaton-Liouville sector,
\eqabegin
 \qpm &=& \overrttwo \left( \qone \pm \qtwo \right)\comma
 \qquad
 \ppm = \overrttwo \left( \pone \pm \ptwo \right)\comma
 \qquad
\left[ \qpm, \ppm \right] = i\comma \\
  \alpm_m &=& \overrttwo \left( \alpha^1_m \pm  \alpha^2_m \right)
\comma \qquad
 \left[ \alpm_m, \alpha^\mp_n \right] = m\deltamn \comma
\eqaend
and assign the degree as
\eqabegin
 \mbox{deg}(\alplus_n ) &=& \mbox{deg}(c_n) =1 \comma
 \qquad
 \mbox{deg}(\alminus_n ) = \mbox{deg}(b_n) =-1 \period
\eqaend
The rest of the mode operators, {\it including the matter part}, are
 defined to carry degree 0.
Then $\dhat$ is decomposed according to the degree
 as
\eqabegin
 \dhat &=& \dhat_0 + \dhat_1 +\dhat_2\comma \\
 \dhat_0 &=& \sum_{n\ne 0} P^+(n)c_{-n}\alminus_n \comma \\
 \dhat_1 &=& \sum_{n.z.m.}:c_{-n}( \alplus_{-m}\alminus_{m+n}
     + \half (m-n)c_{-m}b_{m+n} +\Lf_n) : \comma\\
 \dhat_2 &=& \sum_{n\ne 0} P^-(n)c_{-n}\alplus_n \comma
\eqaend
where $P^\pm(n)$ are given by
\eqabegin
 P^+(n) &=& \overrttwo \left( \pone +\ptwo +2iQn \right)\comma \\
 P^-(n) &=& p^- \period
\eqaend
Notice that, due to the special structure of the background charges,
 $\Pminus(n)$ is independent of $n$ and  further that the $n$
dependence of $\Pplus(n)$ is shifted by 1 unit compared with the
Euclidean case.
It is easy to check that $\dhat_0^2=\dhat_2^2=0$
 holds identically and furthremore,  $\left[ \dhat_0, L_0 \right] =
\left\{ \dhat_0, b_0\right\} =0 $, and similarly for $\dhat_2$. This
 allows us to consider $\dhat_0$- and $\dhat_2$-cohomologies actually
on the entire Fock space $\calF$.  \parsmallskip
Let us summarize, in advance,  the rest of the procedure: Depending on
the conditions on $P^\pm (n)$, we study the $\dhat_0$- or $\dhat_2$-
 cohomology on the Fock space $\calF$ and then restrict it to
the relative cohomology space $\calF_0$.  Then we shall show that
for each case such a cohomology is in one to one correspondence with
the relative $\dhat$-cohomology and explicitly construct the
representatives for the $\dhat$-cohomology.
The final step consists of the construction of $d$-cohomology from
 the $\dhat$-cohomology.  It turns out that except for the single case
where the relative cohomology contains
$\bminusi \pdownvac$, the absolute cohomology can be obtained in the
 simple way:  If $\psi$ is an element of $\dhat$ cohomology,
then $\psi$ and $c_0\psi$ represent the possible $d$ cohomology.
 Therefore in the following, we shall consider explicitly the
$\dhat$ cohomology only and will comment on the $d$ cohomology when
we deal with the special state mentioned above.
\subsubsection{$\Pplus(n) \ne 0\ ({}^\forall n \in {\bf Z}, \
 n \ne 0)$\ Case }
Let us begin with the case in which $\Pplus(n)\ne 0$ for all non-zero
integer $n$.  In this case $\dhat_0$-cohomology is useful since one can
define the operator
\eqabegin
\Kplus &\equiv& \sum_{n\ne 0} {1\over \Pplus(n)} \alplus_{-n} b_n \comma
\eqaend
which satisfies
\eqabegin
 \left\{ \dhat_0, \Kplus\right\} &=& \sum_{n\ne 0}:(nc_{-n} b_n
  + \alplus_{-n} \alminus_n ):+1 \\
 &= & \hat{N}^{dLg} \comma \label{eqn:Nhat}
\eqaend
where $\hat{N}^{dLg}$ is the level counting operator for the
dilaton-Liouville-ghost (dLg) sector.  By a standard argument,
non-trivial $\dhat_0$-cohomology can only be in the sector where
$\hat{N}^{dLg}$ vanishes.  We must also satisfy the $L_0\psi=0$
condition, which in this case reads
\eqabegin
 \pplus\pminus + \half \vecpf^2+\Nhatf -1 &=& 0 \comma
 \label{eqn:Lzerocond}
\eqaend
where $\hat{N}^f$ counts the level for the matter sector.  The states
 satisfying these conditions are indeed $\dhat_0$-non-trivial.  To see
this, look for a state $\Omega$ such that $\dhat_0\Omega$ gives a
state without any non-zero mode exitations in the dLg sector. Because
of the form of $\dhat_0$, $\Omega$ must be of the form
$\sum_{m\ne 0}\left(A_m\alplus_{-m}b_m\right)\Omega_0$, where $\Omega_0$
 is a state without dLg excitations, but obviously such an expression
vanishes. \par
It is evident from Eq.(\ref{eqn:Lzerocond}) that for $\pplus\pminus > 0$
 only a zero-mode excitation is allowed in the matter sector.
  On the other hand, for $\pplus\pminus \le 0$ there
can be matter excitations at non-zero Virasoro levels. As will
 be discussed in Sect.~5, this case will be of
great importance especially when one considers $L\rightarrow \infty$
 limit. \parsmallskip
 We now describe the construction of $\dhat$-cohomology
including the latter case.  Essentially we follow the procedure
described in BMP, but due to the presence of the matter degrees of
freedom, a part of the arguments will have to be modified.  \par
Let $\psi_0$ be a $\dzerohat$ non-trivial state of the form
$F_{-N}\pdownvac$ where $F_{-N}$ is a matter operator at level $N$ and
 $\vec{P} = (\pplus, \pminus, \vecpf)$ .
We have $\dzerohat \psi_0 =0$ and in addition $\dhat_2\psi_0 =0$ as well.
Therefore, $\dhat\psi_0 = \dhat_1\psi_0$ and according to the general
argument of BMP ( see the appendix of \cite{bmp} )  this state must be
$\dhat_0$ exact.  Thus we look for
$\psi_1$ of degree 1 such that $ \dhat_1\psi_0 = -\dzerohat \psi_1$.
Let us apply the operator $\Kplus$ introduced previously.
Then one gets
\eqabegin
 \Kplus\dhat_1\psi_0 &=& -\Kplus \dzerohat\psi_1 \nn\\
 & = &\Biggl[-\left\{\Kplus, \dzerohat\right\}+ \dzerohat\Kplus\Biggr]\psi_1
\nn\\
& = & -\hat{N}^{dLg}\psi_1 + \dzerohat\Kplus\psi_1 \period \label{eqn:psione}
\eqaend
It is instructive to write down the explicit form of $\dhat_1\psi_0$.
It is given by
\eqabegin
 \dhat_1\psi_0 &=& \sum_{n\ge 1} c_{-n}L^f_n F_{-N}\pdownvac \period
\eqaend
Evidently, because of the presence of the matter, this is {\it not}
 an eigenstate of $\hat{N}^{dLg}$ unlike the case treated in BMP.
 Similarly, $\Kplus\dhat_1\psi_0$ is also not an eigenstate.
 Nevertheless, {\it for each term making up such states},
$\hat{N}^{dLg}$ does have positive integral value and this will be
enough to consider the inverse operator $\invNhat$ on such states.
Since the rest of the argument, which is also modified from that in
BMP, is somewhat technical, we shall relegate it to the appendix B. When
 all the dust settles, the result turned out to be  formally identical
to the one obtained in BMP.   Namely, a representative $\psi$ of the
 $\dhat$-cohomology corresponding to a  $\dhat_0$-cohomology
represented by $\psi_0 =F_{-N}\pdownvac$ is given by
\eqabegin
 \psi &=& \sum_{n=0}^\infty (-1)^n (T^+)^n \psi_0 \comma
\label{eqn:dhcohomp}
\eqaend
where the operator $T^+$ is defined by
\eqabegin
 T^+ &\equiv & \invNhat \Kplus \dhat_1 \period \label{eqn:Tplus}
\eqaend
Let us give some examples.  For $F_{-N}= \alpha^i_{-1} \mbox{ and }
 \alpha^i_{-2}$, application of the formula above gives
the following physical states $\ket{\psi_1}$ and $\ket{\psi_2}$:
\eqabegin
 \ket{\psi_1} &=& \Biggl[ \alpha^i_{-1} -{p_f^i\over \Pplus(1)}
   \alpha^+_{-1}\Biggr] \pdownvac \comma\\
 \ket{\psi_2} &=& \Biggl[\alpha^i_{-2} -{2 \over \Pplus(1)}
 \alpha^i_{-1}\alpha^+_{-1} -{p^i_f \over \Pplus(2)}\alpha^+_{-2}
  \nn\\
 & &  + {p^i_f \over \Pplus(1)}\left\{ {1\over \Pplus(1)}
  + {1\over \Pplus(2)}\right\} (\alpha^+_{-1})^2 \Biggr]
 \pdownvac \period
\eqaend
If we were dealing with a string theory, this would describe the
transversality condition.  However, in the dilaton gravity
context, it means that whenever matter fields are present
 they necessarily induce excitations in the dilaton-gravity sector.
\subsubsection{$\pminus \ne 0$\quad Case }
In this case, the relevant tool is the $\dhat_2$-cohomology.
 The procedure is entirely similar to the previous case, except
 for the use of $K^-\equiv (1/p^-)\sum_{n\ne 0}\alminus_{-n}b_n$
 in place of $K^+$.  $L_0\psi=0$ condition is the same as
 given in (\ref{eqn:Lzerocond}) and
we obtain the representative for $\dhat$-cohomology corresponding
to Eqs.(\ref{eqn:dhcohomp}), (\ref{eqn:Tplus}) as
\eqabegin
 \psi &=& \sum_{n=0}^\infty (-1)^n (T^-)^n \psi_0 \comma
 \label{eqn:dhcohomm} \\
 T^- &\equiv & \invNhat \Kminus \dhat_1 \period \label{eqn:Tminus}
\eqaend
%
\subsubsection{$\Pplus(r)=0, \quad \pminus =0$\quad Case}
Now we consider the remaining case where $\Pplus(r)=0$ for some integer
$r$ and at the same time $\pminus=0$.  Physical states for this case
have come to be called \lq\lq discrete states".  We look at the
$\dhat_0$-cohomology and define the operator $K^+_r$ analogously to
$K^+$ except without the term involving $\Pplus(r)$. That is
\eqabegin
 \Kplus_r &\equiv& \sum_{n\ne 0, r} {1\over \Pplus(n)}
 \alplus_{-n} b_n \period
\eqaend
This operator satisfies the relation
 $\left\{ \dhat_0, K^+_r \right\}=\hat{N}_r^{dLg}$, where
$\hat{N}_r^{dLg}$ is the level-counting operator for the dLg sector
 this time excluding the $r$-th level,  and it must vanish in order for
$\dhat_0$-cohomology to be non-trivial. $L_0\psi=0$ condition now reads
\eqabegin
 \half\vecpf^2 + \Nhatf + \Nhatdlg -1 &=&0 \period
\eqaend
Since $\vecpf^2$ and $\Nhatf$ are non-negative, we see that $\Nhatdlg$
 can either be 1 or 0.  Together with the vanishing of $\hat{N}_r^{dLg}$
this means that $r = \pm 1$ and only a level 1 excitation is possibly
 allowed in the dLg sector.   We now study these cases separately.
\paragraph{$r=1$ case: }
When there is an excitation at level 1 in the dLg sector, $\vecpf=0$
 and we find two types of $\dhat_0$-cohomology represented by
 the following states:
\eqabegin
  & &\left(\Aplus\alplusi + \vec{A}\cdot\vecali\right)\pdownvac
 \quad \mbox{(ghost number 0)}\comma \\
  & & c_{-1} \pdownvac \quad \mbox{(ghost number 1)}\comma
\eqaend
where $A_+$ and $\vec{A}$ are arbitrary coefficients.  These states
 are easily checked to be $\dhat$ non-trivial as well.
For the latter state, it is instructive to compare with the usual
bosonic string case with $Q=0$.
  In that case all the momenta vanish and we have
$c_{-1}\mid \vec{0} >$. This state however is $d$-exact.  Indeed
\eqabegin
 d c_0b_{-1}\mid \vec{0}>_\downarrow &=& b_0M c_0b_{-1}
\mid \vec{0}>_\downarrow
 = c_{-1}\mid \vec{0} > \period
\eqaend
This phenomenon is allowed because BMP Theorem 4.2 breaks down.
Namely, in this particular configuration of the momenta, there exists
 another $\dhat$-non-trivial state $b_{-1}\mid \vec{0} >$ at
ghost number -1 and the argument of Theorem 4.2 which assumes non-
existence of cohomologies separated by 2 units of ghost number
is no longer valid.  \par
 Now for  $\Nhatdlg =0$, we simply have $\pdownvac$ as representing
non-trivial $\dzerohat$ and $\dhat$ cohomology, where $\vecpf^2 =2$.
\paragraph{$r=-1$ case: }
Going through an analysis similar to the previous case using $K^+_{-1}$,
 we find the following two non-trivial $\dzerohat$ cohomology
representatives:
\eqabegin
 & & \left( \Aminus \alminusi + \vec{A}\cdot\vecali \right)\pdownvac
  \comma\\
 & & b_{-1}\pdownvac \period
\eqaend
It is easy to check that these states represent $\dhat$ cohomologies
 as well. \par
 Finally we must make a comment on the absolute cohomology
which arises from $\psi \equiv b_{-1}\pdownvac$. According to the
general argument of BMP, we have, besides $\psi$ itself, the
second member $c_0\psi + \chi$,  where $\chi$ is determined by the
equation  $M\psi = \dhat \chi$.  Explicitly, this reads

\eqabegin
 M\psi &=& 2c_{-1}c_1 b_{-1}\pdownvac
  = 2c_{-1}\pdownvac \nn\\
  &=& \dhat {2\over \Pplus(1)}\alplusi\pdownvac \period
\eqaend
Thus the second member of the absolute cohomology takes the form
\eqabegin
 & & \left( c_0b_{-1} +{2\over \Pplus(1)}\alplusi\right)\pdownvac
 \period
\eqaend
As mentioned previously, this is the only exception in which the
second member of the absolute cohomology is not given simply by
appending the $c_0$ operator to a representative of $\dhat$-cohomology.
\subsection{Physical Operators}
Having obtained all the BRST non-trivial states, we now wish to
discuss the corresponding physical operators.  It is well-known that
 in the {\it Euclidean plane coordinate} formulation a state
$\ket{\psi}$ corresponding to an operator $\Psi(z, \bar{z})$ is
given by
\eqabegin
 \lim_{z \rightarrow 0 \atop \bar{z}\rightarrow 0}
 \Psi(z, \bar{z})\ket{0} &=& \ket{\psi} \comma \label{eqn:opst}
\eqaend
where $\ket{0}$ is the $sl(2)$ invariant vacuum.  However, in the
{\it Minkowski cylinder coordinate} we cannot directly use this
procedure.    Thus we shall first
develop a useful machinery which allows us to convert between
these two types of formulations and then try to make use of
 the simple correspondence (\ref{eqn:opst}).  \parsmallskip
Consider in Minkowski space the left conformal transformation of a
scalar field $\phi$ with a background charge $Q$.  It is given by
\eqabegin
 \left[T_\epsilon, \phi(\xplus)\right]
 &=& {1\over i}\left( \epsilon(\xplus)\delplus\phi(\xplus)
 +Q\delplus\epsilon(\xplus) \right) \comma
\eqaend
where
\eqabegin
 T_\epsilon &=& \sum \epsilon_{-n}L^M_n \comma\\
 \epsilon(\xplus) &\equiv& \sum \epsilon_{-n} \e^{in\xplus} \comma\\
 \left[ L^M_n, \phi(\xplus)\right] &=& \e^{in \xplus}
 \left( {1\over i}\delplus\phi + Qn \right) \period
\eqaend
This implies that for
  a finite tranformation (including the right-transformation) given by
$y^+ =y^+(x^+),\ y^- =y^-(x^-)$, the field $\phi$ transforms as
\eqabegin
 \phi(y^+) &=& \phi(\xplus) -Q\ln\left({d y^+ \over d\xplus}\right)
-Q\ln\left({d y^- \over d\xminus}\right) \period
\eqaend
Applying this result to the case of interest, namely
\eqabegin
 y^+&=& z = \e^{i\xplus} \comma\\
 y^-&=& \bar{z}= \e^{i\xminus} \comma
\eqaend
we immediately get
\eqabegin
\phi^{EP} (z,\bar{z}) &=& \phi^M (\xplus,\xminus)
-iQ(\xplus+\xminus) -2Q\ln(i) \comma
\eqaend
where we have supplemented the superscript EP and M to distinguish the
Euclidean-Plane and Minkowski fields.  By writing out the Fourier
components,
this means the following identifications:
\eqabegin
 q^{EP} &=& q^M -iQ\pi  \comma\\
 p^{EP} &=& p^M -iQ \comma\\
 \alpha_n^{EP} &=& \alpha_n^M \period
\eqaend
An alternative more useful way of effecting this transformation is to
agree to use {\it the same mode operators} and regard it
as a similarity transformation.  Specifically,
\eqabegin
 \calU \phi^{EP}\calUinv &=&  \phi^M  \comma\\
 \calU &=& \e^{Qq}\e^{Q\pi p} \period
\eqaend
It is not difficult to check explicitly that this operation
correctly converts the Virasoro generators. Namely,
\eqabegin
 L^{EP}_n &=& \half :\sum_m \alpha_{n-m}\alpha_m :
 + iQ(n+1)\alpha_n \comma\\
\calU L^{EP}_n \calUinv &=&  L^M_n  \comma\\
 L^M_n &=& \half :\sum_m \alpha_{n-m}\alpha_m :
 + iQn\alpha_n + {Q^2\over 2}\delta_{n,0} \period
\eqaend
It is obvious from this relation that $L^{EP}_n$ and $L^M_n$
both satisfy the standard form of the Virasoro algebra. \parsmallskip
Now the operator $\Psi(\xplus, \xminus)_M$ corresponding to a given
state $\ket{\psi}_M$ in Minkowski space can be found through the
following procedure:
 First get the corresponding Euclidean state $\ket{\psi}_E$ by
$\ket{\psi}_E = \calUinv \ket{\psi}_M$.  Then by the usual correspondence
 find the Euclidean operator $\Psi(z,\bar{z})_E$.  Finally, it is
converted to the Minkowski operator by
$\Psi(\xplus,\xminus)_M = \calU \Psi(z,\bar{z})_E \calUinv$. Since
 the conversion is done by a similarity transformation, conformal
 and BRST properties of the states and the fields are guaranteed to
 be preserved. \parsmallskip
In particular, all the physical operators corresponding to the
physical states obtained in the previous subsection can easily be
constructed.  One thing we must be careful about in this procedure,
 however, is
that the normal  ordering is properly defined for the independent
fields $\phi_i$ and not for $\psi$ and $\chi$.
Let us give an example at level 1 to illustrate this point. Consider
 a state of the  form $(A_+ \alpha^+_{-1} +
 \vec{A}\cdot\vec{\alpha}^f_{-1})\pdownvac$, which is a
 physical state for $\pone = \ptwo = -iQ,\ \vecpf =0$.  The corresponding
 operator invariant under the left BRST transformation is given
 by
\eqabegin
 & & c\Biggl[A_+ \left\{ \left(\delplus :\e^{2Q\phione}:\right)
 :\e^{-2Q\phitwo}:
 -:\e^{2Q\phione}:\left(\delplus :\e^{-2Q\phitwo}:\right)
\right\}  \nn\\
 & &
 + \vec{A}\cdot\delplus \vec{f}:\e^{2Q\phione}: :\e^{-2Q\phitwo}:
 \Biggr] \comma
\eqaend
where the symmetric normal ordering must be adopted.
\newpage
\section{Discussions} \sectionnumbering
Starting from a definite model of dilaton gravity given by the action
Eq.(\ref{eqn:ht}), we have succeeded in its operator quantization
 and analyzed the physical states and  operators of the theory in
 the BRST framework.  Our emphasis throughout is  the rigor of the
 analysis,  trying to avoid as much as possible any explicit or implicit
  assumptions and preconceptions ,
which often obscure the validity of the results  obtained.
This is particularly important for quantum  gravity since it
  possesses many features quite distinct from ordinary quantum field
theory.    \parsmallskip
The work we have performed in this article constitutes the first stage
of our intended investigation.  The task of extracting  physical
 consequences of the model remains to be undertaken.
 We now wish to list and analyze in some detail the nature of the
 problems that lie ahead,  which would be invariably encountered
 in any attempt for exact treatment of models of quantum gravity.
 \parsmallskip
With the knowledge of the physical states and operators in hand,
the obvious next step is to calculate the \lq\lq matrix elements"
 and interprete them physically.  There are a number of closely
 intertwined problems, both technical and conceptual,
 to be solved at this stage.  The issue centers around the
interpretation of the \lq\lq wave function", namely the difficulty of
 interpreting it as a probability amplitude.  Perhaps the best attitude
 toward this problem is to try to stick rigorously to the
principle of quantum mechanics and see what that leads to. \parsmallskip
 In our setting, the first specific problem to be solved is what inner
product to be introduced in the space of states.  The canonical
commutation relation alone does not tell us the answer since we can have
infinitely many different representions of the Heisenberg algebra.
 In an ordinary quantum field theory, this problem is settled by
demanding appropriate hermiticity property for the basic operators
 so that the physical observables have real eigenvalues and that
 the Hamiltonian is bounded from below.  Clearly, this criterion cannot
 be applied directly in quantum gravity.  Let us nevertheless try to see
 how far we can go along this line.  \parsmallskip
First we show that possible hermiticity assignments for the modes are
 severely restricted once we adopt the usual assignments for the ghosts
$b$ and $c$,  namely
\eqabegin
 c^\dagger_n &=& c_{-n}, \qquad b^\dagger_n =b_{-n} \period
\eqaend
The argument goes as follows:
 Consider a physical matrix element of a BRST invariant operator
$\cal{O}$
\eqabegin
 & & \left(\ket{phys}, {\cal O } \ket{ phys' }\right) \period
\eqaend
Since this  should be independent of the choice of the representative of
 the cohomology, we must have
\eqabegin
0 &=& \left(\ket{phys}, {\cal O } d \ket{\ast}\right) \nn\\
 &=& \left(d^\dagger \ket{phys}, {\cal O} \ket{\ast}\right) \comma
\eqaend
where $d$ is the BRST operator.  This means that $d^\dagger$ must always
annihilate physical states and
hence we must require $d^\dagger = \mbox{const.}\times d$.  But since
 the part of $d$ consisting of ghosts alone is hermitian, the constant
 above can only be unity.  This in turn dictates the hermiticity
property of the virasoro generator to be
\eqabegin
 L_n^\dagger &=& L_{-n} \period
\eqaend
In general this does not fix the hermiticity property of $\alpha_n$
completely,  but for the dilaton-Liouville sector it does. The
 key is the presence of the background charge term
 $iQn(\alpha^1_n -\alpha^2_n)$  in $L^{dL}_n$, which is linear in the
 oscillator.  Since $Q$ for our model is real, this leads to
$(\alpha^i_n)^\dagger = \alpha^i_{-n}$ for $n\ne 0$.  Once the assignment
 for the non-zero mode is fixed as above, the zero mode $\alpha_0$
( and its conjugate ) should be taken to be hermitian since
$(\alpha_0\alpha_n)^\dagger = \alpha_0\alpha_{-n}$ must hold. \par
This however still does not settle the question of the inner product.
As was first discussed in detail in \cite{afio} and subsequently
 applied to the case of gravity in \cite{afko}, there are
essentially two different realizations of the Heisenberg algebra with
 the hermiticity assignments deduced above.  The one that was argued
to be relevant to quantum gravity is associated with the inner
 product with {\it indefinite norm}, for which hermitian operators may
have imaginary eigenvalues.  We believe that this is a point of utmost
importance which captures the very characteristic of gravity distinct
 from usual field theories. \par
In fact we can already see its relevance  in our analysis of physical
states.  First, we have seen that for the discrete states with
$\Pplus(r)=0$,  the value of the zero mode $\pplus$ is imaginary.
Furthermore, for states involving matter excitations, it may  be
 more relevant.  Recall the $L_0 \psi =0$ condition (\ref{eqn:Lzerocond})
\eqabegin
 \half(\pone^2-\ptwo^2) + \half\vec{p}_f^2 + \hat{N}^f -1 &=& 0
\period
\eqaend
As will be discussed shortly, states containing matter fields
 with finite momentum in the limit $L\rightarrow\infty$ must
 have matter excitations at arbitrary high Virasoro levels.  From the
condition above, this is possible only if $\pone^2-\ptwo^2 $ can take
arbitrary large negative values.  There are several ways of realizing
this condition, including the one with imaginary values for $p_i$'s.
 Afterall, the equation above should be expressing the energy balance
between the matter and the dilaton-
Liouville sectors, characteristic of a theory of gravity. Thus at the
least, we must carefully examine the appropriate choice of the
inner product in order to compute the matrix elements and to
make physical interpretation of them, including the question of
\lq\lq unitarity" of the theory. \parsmallskip
Deeply linked with the  problem above is the question of how to
extract physics.  To see what sort of averaged spacetime configuration
is associated with a given physical state, we must evaluate
expectation  values of some appropriate operators in such a state.
BRST invariant operators are certainly preferred, but it is not an easy
task to find such operators of direct physical significance.  Another
possibility is to devise a way to completely fix the gauge freedom
 within the conformal gauge and try to evaluate an operator, such
 as the metric, which allows for more direct interpretation.
Preliminary investigation indicates that this attempt also requires some
ingenuity to be successful.
\parsmallskip
At this juncture, let us make a brief remark on an  important
point which should be kept in mind when one tries to deal with
a \lq\lq quantum black hole" in {\it two dimensions}.
 It has to do with the notion of asymptotic flatness and that of the
\lq\lq mass" of the black hole.  In semi-classical treatment, one
 first identifies a classical black hole configuration and then
finds the Hawking radiation emitted into the asymptotically flat region
 of such a background, with subsequent modification of the metric
 due to the radiation itself, \ie the back reaction.  This
sequential procedure can no longer be applied in an exact treatment
 such as the one we have been pursuing.  The metric, the dilaton and the
matter degrees of freedom are inherently intertwined and especially
 in one spatial dimension, where the radiation does not disperse, there
cannot be an asymptotically flat region.  This is already indicated
 in the form of a physical state with matter excitation discussed
 in the previous section.  Such a state is necessarily accompanied by
 excitations in the dilaton-Liouville sector.  This would make the
identification of the black hole \lq\lq mass" even more difficult.
\parsmallskip
Finally, let us discuss the question of the large $L$ limit.
The problems described so far can be posed already for the case
 of the universe with finite size $L$.  If one however wishes to study
what happens in the $L\rightarrow\infty$ limit, one must face a
further technical problem.  As was remarked at the beginning of Sect.~4,
 the structure of the physical states of  our model is very similar
to that of a critical closed string theory.  However, its interpretation
is quite different.  The $p$'s should be interpreted only as a set of
variables describing the zero modes of the theory and not as
physical momenta.  The actual momenta are discretized in the units of
 $1/L$ and are associated with higher Virasoro levels.  Consequently,
in the limit of large $L$,  states carrying finite physical momenta
 correspond in string language to excited states at arbitrary
 high Virasoro levels.  Moreover, as we have seen already in the
analysis of classical solutions,  large degeneracy is inevitable in this
 \lq\lq infrared" limit.  Bumpy structures which may be seen in  a
finite universe can be artifacts to be flattened away as $L$ is
taken to infinity.  Although we believe that a lot of physics should be
extractable for finite universe, we must keep this difficulty in mind
 and try to develop some clever means to deal with it. \parsmallskip
Although undoubtedly quite challenging as they are, the problems
 described above are all extremely intriguing  and worth pursuing.
   They are presently  under  investigation  and we
hope to be able to report our progress elswhere. \parsmallskip
While preparing the manuscript, we received a preprint \cite{VV}
 which deals with a model similar to ours with $24$ matter scalars.
The treatment of the model, especially the boundary condition, is
 however different from ours.
\parbigskipn
\parbigskipn
{\Large\bf Acknowledgment} \parbigskipn
Y.K. would like to acknowledge stimulating discussions with
K. Hamada, M. Kato and H. Kodama,  while S.H. is grateful to H. Ishikawa
 for helpful advices.  The research of Y.K. is supported in part
by the Grant in Aid for Scientific Research from the Ministry of Education,
Science and Culture No.04245208 and No.04640283.
\noindent\parbigskipn\parbigskipn
{\Large\bf Appendix A} \parbigskipn
In this appendix, we list some useful Poisson brackets needed for
the calculation of $\left\{\chi(x), AB(y)\right\}$ discussed in
section 3.1.  \par
 As we remarked in the text, one needs to carefully take into account
the fact that the quatntity $\alpha$, which specifies the boundary
conditions for $A(\xplus)$ and $B(\xminus)$, is a function of $P_\psi$.
{}From its form $\alpha =\exp(P_\psi /2)$, we derive
\eqabegin
 \left\{ \chi(x), \alpha \right\}
   &=& \left\{ Q_\chi, \e^{P_\psi/2} \right\} = \half \alpha \comma
  \nn\\
  \left\{ \chi(x), C(\alpha) \right\}
 &=& \half\alpha{\del\over \del \alpha} C(\alpha)
 = -{1\over 4} C(\alpha)^2\left( \sqrt{\alpha}
 +{1\over \sqrt{\alpha}}\right)\comma \nn\\
 \left\{ \chi(x), E_\alpha(\sigma_y -\sigma') \right\}
 &=& {1\over 4} \epsilon(\sigma_y -\sigma')
   E_\alpha(\sigma_y -\sigma')\comma \nn\\
 \left\{ \chi(x), E_{1/\alpha}(\sigma_y -\sigma'') \right\}
 &=& -{1\over 4} \epsilon(\sigma_y -\sigma'')
   E_{1/\alpha}(\sigma_y -\sigma'') \period\nn
\eqaend
Using these formulae as well as Eqs.(\ref{eqn:chipsi}) $\sim$
(\ref{eqn:psichi}) in the text,
we obtain the following basic Poisson brackets evaluated at equal time:
\eqabegin
 \left\{ \chi(\sigma_x), A(\sigma_y)\right\}_{ET}
 &=& {\sigma_x \over 4\pi}A(\sigma_y)
  -\half C(\alpha)A(\sigma_x) E_{1/\alpha}(\sigma_x -\sigma_y)
    -{1\over 4}A(\sigma_y)\epsilon(\sigma_x-\sigma_y) \comma\nn\\
 \left\{ \chi(\sigma_x), \dot{A}(\sigma_y)\right\}_{ET}
 &=& {\sigma_x \over 4\pi}\dot{A}(\sigma_y)
    -{1\over 4}\dot{A}(\sigma_y)\epsilon(\sigma_x-\sigma_y)\comma \nn\\
 \left\{ \dot{\chi}(\sigma_x), A(\sigma_y)\right\}_{ET}
 &=& -\half C(\alpha) \dot{A}(\sigma_x)
        E_{1/\alpha}(\sigma_x -\sigma_y) \comma\nn\\
 \left\{ \dot{\chi}(\sigma_x), \dot{A}(\sigma_y)\right\}_{ET}
 &=& -\half \dot{A}(\sigma_x)\delta(\sigma_x -\sigma_y) \comma\nn\\
 \left\{ \chi(\sigma_x), B(\sigma_y)\right\}_{ET}
 &=& -{\sigma_x \over 4\pi}B(\sigma_y)
  -\half C(\alpha) E_{1/\alpha}(\sigma_y -\sigma_x)B(\sigma_x)
    +{1\over 4}B(\sigma_y)\epsilon(\sigma_x-\sigma_y) \comma\nn\\
 \left\{ \chi(\sigma_x), \dot{B}(\sigma_y)\right\}_{ET}
 &=& -{\sigma_x \over 4\pi}\dot{B}(\sigma_y)
    +{1\over 4}\dot{B}(\sigma_y)\epsilon(\sigma_x-\sigma_y) \comma\nn\\
 \left\{ \dot{\chi}(\sigma_x), B(\sigma_y)\right\}_{ET}
 &=& -\half C(\alpha) E_{1/\alpha}(\sigma_y -\sigma_x)\dot{B}
 (\sigma_x) \comma\nn\\
 \left\{ \dot{\chi}(\sigma_x), \dot{B}(\sigma_y)\right\}_{ET}
 &=& -\half \dot{B}(\sigma_x)\delta(\sigma_x -\sigma_y) \period\nn
\eqaend
Combining these equations it is now straightforward to get
\eqabegin
 \left\{ \chi(\sigma_x), AB(\sigma_y) \right\}_{ET}
 &=& -\half C(\alpha) \left[ A(\sigma_x)
     E_{1/\alpha}(\sigma_x -\sigma_y)B(\sigma_y)
  + A(\sigma_y) E_{1/\alpha}(\sigma_y -\sigma_x)B(\sigma_x)
   \right] \comma\nn\\
 \left\{ \dot{\chi}(\sigma_x), AB(\sigma_y) \right\}_{ET}
 &=& -\half C(\alpha) \left[ \dot{A}(\sigma_x)
     E_{1/\alpha}(\sigma_x -\sigma_y)B(\sigma_y)
  + A(\sigma_y) E_{1/\alpha}(\sigma_y -\sigma_x)\dot{B}(\sigma_x)
   \right] \comma\nn\\
 \left\{ \chi(\sigma_x), \del_t(AB)(\sigma_y) \right\}_{ET}
 &=& -\half C(\alpha) \left[ A(\sigma_x)
     E_{1/\alpha}(\sigma_x -\sigma_y)\dot{B}(\sigma_y)
  + \dot{A}(\sigma_y) E_{1/\alpha}(\sigma_y -\sigma_x)B(\sigma_x)
   \right] \comma\nn\\
 \left\{ \dot{\chi}(\sigma_x), \del_t(AB)(\sigma_y) \right\}_{ET}
 &=& -\half\delta(\sigma_x-\sigma_y)\del_t(AB)(\sigma_y)\nn\\
 & &
-\half C(\alpha) \left[ \dot{A}(\sigma_x)
     E_{1/\alpha}(\sigma_x -\sigma_y)\dot{B}(\sigma_y)
  + \dot{A}(\sigma_y) E_{1/\alpha}(\sigma_y -\sigma_x)\dot{B}(\sigma_x)
   \right] \period\nn
\eqaend
Notice that these expressions are all symmetric with respect to the
 interchange  $x\leftrightarrow y$.  Therefore the combination
$$
 \left\{ \chi(\sigma_x), AB(\sigma_y)\right\}_{ET}
- \left\{ \chi(\sigma_y), AB(\sigma_x)\right\}_{ET}
 $$
and the time derivatives thereof ( at most one each for $x$ and $y$ )
 all vanish. \parbigskipn\parbigskipn
{\Large\bf Appendix B} \parbigskipn
In this appendix, we shall supply the arguments needed to justify
 the result (\ref {eqn:dhcohomp}),(\ref{eqn:Tplus}) in subsection 4.1 of the
text.  \par
First we prove that the solution of (\ref{eqn:psione}) is given by
\eqabegin
 \psi_1 &=& -\invNhat \Kplus\dhat_1\psi_0 \period\nn
\eqaend
For this to be correct, we must show that $\Kplus\psi_1=0$.  Noting
that $\Kplus$ is a level zero operator in the dLg sector
 and hence commutes with $\invNhat$,  we get
\eqabegin
 \Kplus \psi_1 &=& -\Kplus\invNhat \Kplus \dhat_1\psi_0 \nn\\
  &=& -\invNhat (\Kplus)^2 \dhat_1\psi_0 \period\nn
\eqaend
This indeed vanishes because $(\Kplus)^2 =0$. \par
Next we prove the important property that $\dhat_2\psi_1 =0$ holds.
 Since $\dhat_2$ and $\invNhat$ again commute, we have
\eqabegin
 \dhat_2\psi_1 &=& -\dhat_2\invNhat \Kplus \dhat_1\psi_0 \nn\\
 &=& -\invNhat \dhat_2\Kplus\dhat_1\psi_0 \nn\\
 &=& -\invNhat  \left\{ \dhat_2, \Kplus\right\}
 \dhat_1\psi_0 + \invNhat \Kplus \dhat_2\dhat_1\psi_0 \nn\\
 &=& -\invNhat \sum_{n\ne 0}{\pminus \over
 \Pplus(n)} \alplus_{-n}\alplus_n \dhat_1\psi_0
 -\invNhat \Kplus \dhat_1\dhat_2\psi_0 \nn\\
 &=& 0 \comma\nn
\eqaend
where we used the fact that
 $\dhat_1\psi_0$ does not contain any $\alminus_{-n}$ and hence gets
annihiliated by $\alplus_n$.  \par
We now go to the next step.  Form $\psi_0+\psi_1$ and act $\dhat$ on
it.  Because $\dhat_2\psi_1=0$, we get $ \dhat(\psi_0+\psi_1) =
 \dhat_1\psi_1$.
Again from the genral argument this state must be $\dzerohat$-exact
 and we look for a state $\psi_2$ of degree 2 such that
$ \dhat_1\psi_1 = -\dzerohat\psi_2 $.
It is clear that we can repeat the previous procedure and find a
solution
$$ \psi_2 = -\invNhat \Kplus\dhat_1\psi_1 .$$
In order for this process to terminate, we must show that there exists
 the maximum degree.  This is easy to prove once we note that all the
degrees are carried only by the
 oscillators in the dLg sector. ( In the construction above, the degree
 is carried solely by $\alplus_{-n}$'s.) Thus every time the degree
is raised
by 1, the Virasoro level of the dLg sector is increased at least
 by one unit.  But since the total Virasoro level, including the
matter sector, must stay constant, namely at $N$, it means that
at every step, the level in the matter sector is decreased by at least
one unit. Thus, after at most $N$ steps, the process terminates.
In this way we have the general solution (\ref{eqn:dhcohomp}),(\ref{eqn:Tplus})
 for the relative cohomology.

\end{document}